# *Controlled Electrochemical Intercalation of Graphene/h-BN van der Waals Heterostructures*

S.Y. Frank Zhao[†], Giselle A. Elbaz[‡], D. Kwabena Bediako[†], Cyndia Yu[†], Dmitri K. Efetov[§], Yinsheng Guo[‡], Jayakanth Ravichandran[†], Kyung-Ah Min[⊥], Suklyun Hong[⊥], Takashi Taniguchi[#], Kenji Watanabe[#], Louis E. Brus[‡], Xavier Roy[‡], and Philip Kim[†*]

[†] Department of Physics, Harvard University, Cambridge, Massachusetts 02138, United States

[‡] Department of Chemistry, Columbia University, New York, NY 10027, United States

[§] Department of Physics, Columbia University, New York, NY 10027, United States

[⊥] Department of Physics and Graphene Research Institute, Sejong University, Seoul 143-747, South Korea

[#] National Institute for Materials Science, 1-1 Namiki Tsukuba, Ibaraki 305-0044, Japan

**ABSTRACT:** Electrochemical intercalation is a powerful method for tuning the electronic properties of layered solids. In this work, we report an electrochemical strategy to controllably intercalate lithium ions into a series of van der Waals (vdW) heterostructures built by sandwiching graphene between hexagonal boron nitride (*h*-BN). We demonstrate that encapsulating graphene with *h*-BN eliminates parasitic surface side reactions while simultaneously creating a new hetero-interface that permits intercalation between the atomically thin layers. To monitor the electrochemical process, we employ the Hall effect to precisely monitor the intercalation reaction. We also simultaneously probe the spectroscopic and electrical transport properties of the resulting intercalation compounds at different stages of intercalation. We achieve the highest carrier density $> 5 \times 10^{13}$ cm$^2$ with mobility $> 10^3$ cm$^2$/Vs in the most heavily intercalated samples, where Shubnikov-de Haas quantum oscillations are observed at low temperatures. These results set the stage for further studies that employ intercalation in modifying properties of vdW heterostructures.

Graphite intercalation compounds (GICs) exhibit a variety of interesting properties that differ significantly from semimetal graphite [1]. For example, CaC$_6$ and YbC$_6$ display superconductivity [2], while Li$_{0.25}$Eu$_{1.95}$C$_6$ and EuC$_6$ exhibit ferro- and antiferromagnetic ordering, respectively [3]. Intercalation compounds also represent technologically significant materials. LiC$_6$ is the prototypical anode material in Li ion batteries. By analogy to these bulk graphite intercalation compounds, the intercalation of few-layer graphene has also been realized [4, 5, 6]. Upon intercalation of Li, the optical properties of few-layer graphene crystals (with thicknesses down to 1 nm) change significantly [4], Ca-intercalated few-layer-graphene is superconducting [5], and FeCl$_3$ intercalated bilayer graphene showed a hint of ferromagnetism [6]. In addition, there have been theoretical predictions that heavy doping and proximity induced spin-orbit coupling from certain intercalants may induce exotic electronic properties in the graphene channel [7].

Recently, it was demonstrated that one can stack different van der Waals (vdW) atomic layers to form vdW heterostructures, creating a new generation of few-atomic-layer functional heterostructures with emergent properties [8, 9]. In particular, graphene encapsulated by *h*-BN, a layered insulator, forms a vdW heterostructure where the 2-dimensional (2D) graphene channel is well isolated from the environment [8]. As in intercalation compounds of bulk vdW materials, the intercalation of vdW heterostructures may create a new generation of functional heterostructures with emergent properties. Furthermore, the use of *h*-BN protecting layers may enable the formation of stable intercalation compounds that differ significantly from the bulk intercalation compound due the presence of two dissimilar surfaces at the hetero-interface [10].

Compared to traditional intercalation methods for van der Waals materials, these synthetic *h*-BN/graphene vdW heterostructures present several challenges for intercalation. For example, bulk alkali metal intercalated vdW crystals are chemically highly unstable, prohibiting subsequent exfoliation into few-atomic-layer intercalated nanocrystals. Conversely, conventional chemical intercalation methods involve highly reactive reagents and high temperatures [1], often incompatible with microfabrication procedures for electronic device characterization. Another challenge associated with electrochemical intercalation of atomically thin vdW heterostructures stems from the difficulties in measuring the sub-picoampere electrochemical currents produced from atomically thin van der Waals structures having micron-size lateral dimensions. Such a small current can easily be dominated by current contributions from parasitic reactions occurring in the electrolyte, precluding the use of standard electrochemistry techniques such as cyclic voltammetry.

Our strategy to overcome these hurdles was to employ an electrochemical technique on a pre-fabricated mesoscopic electrical device, thus replacing conventional molten metal reagents with a relatively inert electrolyte, and using the applied bias to deliver a controllable driving force. We demonstrate (i) that *h*-BN is an effective passivation layer for 2D devices with respect to electrochemical degradation; (ii) the formation of a prototype heterostructure intercalation compound and show for the first time the insertion of Li ions into the interface between single-layer graphene and *h*-BN crystals; (iii) the use of the Hall effect to monitor the progress of intercalation as a function of applied bias, rather than standard voltammetry methods [11]. Our approach is not restricted only to graphene/*h*-BN heterostructures, or only to the intercalation of Li; it can be generalized to a wide range of heterostructures, opening the field to a new system of intercalation compounds.

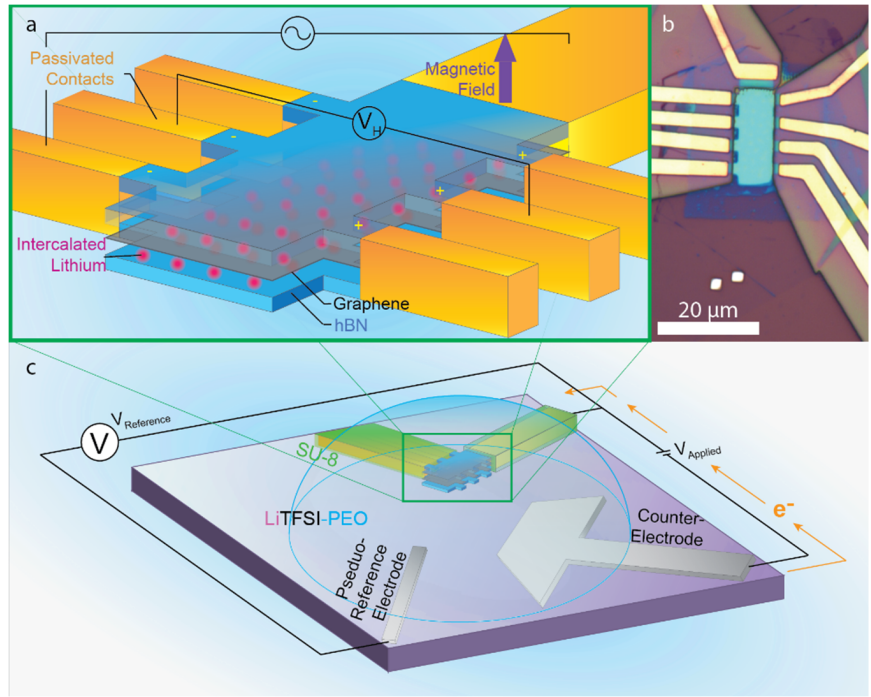

*Figure 1: (a) Schematic of the Hall bar device. A hBN/Graphene/hBN heterostructure is patterned into a Hall bar with edge Cr/Pd/Au (2/15/60 nm) contacts. A channel is left open at the entrance of the Hall bar to allow interactions with the electrolyte. An AC current is applied across the device in a 0.5T magnetic field, inducing a transverse Hall voltage across the device. The Hall voltage is used to monitor the intercalation reaction. (b) Optical micrograph of a representative heterostructure device before electrolyte deposition. Note that all electrodes are covered with SU-8 photoresist, leaving only an edge of graphene in contact with the electrolyte. (c) The solid polymer LiTFSI-PEO electrolyte is dropcast over the Hall bar device (working electrode) shown in (a) as well as a Pt pseudo-reference electrode and counter electrode. To drive the electrochemical reaction, a voltage is applied between the counter electrode and working electrode, and the intercalation voltage is measured versus a Pt pseudo-reference.*

In our experiment, we use mechanical exfoliation followed by van der Waals dry assembly techniques [12] to fabricate vdW heterostructures, using monolayer (1LG) or bilayer (2LG) graphene sandwiched between ~30 nm thick *h*-BN crystals. The vdW stacks are then shaped into Hall-bar geometries, with only the well-defined graphene edges in the vdW stack exposed to the electrolyte. Intercalation is thus only allowed from the edge of the sample (fig. 1b). Using this technique, we can unambiguously and directly observe the reversible doping of graphene as Li ions intercalate and deintercalate the *h*-BN/graphene interface, despite any side reactions that take place at electrolyte-exposed conducting surfaces. Figure 1a-c depicts the basic design of our electrochemical cell. Here, we use the graphene channel as a working electrode. The graphene channel itself is electrically contacted by gold electrodes using the edge contact method [12]. All gold contacts and wires in our devices are covered with a passivating SU-8 layer that is both electrochemically inert and electrically insulating. This simultaneously protects the Au from corrosive reactions occurring at high voltage and limits the number of side reactions occurring in the cell. The device includes a Pt counter electrode and a Pt pseudo-reference electrode. We cover the whole device with a solid electrolyte composed of lithium bis(trifluoromethane)sulfonimide (LiTFSI) suspended in a polyethylene oxide (PEO) matrix. (See detailed methods in SI)

To control the intercalation progress, we monitor intercalation in real time using Hall effect measurements (with a small applied magnetic field of 0.5 T), simultaneously probing the electrical transport properties and the charge carrier density of the crystals as Li ions are inserted [11]. Using this technique, we can unambiguously observe the reversible doping of graphene as Li ions intercalate and deintercalate the *h*-BN/graphene interface. Figure 1b shows an optical microscope image of a typical device used for the experiment. To facilitate magneto-transport measurements, the graphene heterostructure is patterned into a Hall bar geometry. Note that the standard Hall bar geometry is modified slightly, with the source contact at the end of the Hall bar split into two on the far side of the device, so that the corresponding etched edge can be exposed directly to the electrolyte.

Figure 2 shows the resistivity and estimated carrier density obtained from Hall measurement for two devices built from 1LG and 2LG sandwiched between *h*-BN. In both cases, while the electrode potential is swept towards increasingly negative values at 325 K (i.e. towards more reducing potential), the graphene channel carrier density increases linearly with the applied voltage, while the resistance of the sample decreases, consistent with electrostatic gating of the graphene crystal, through the *h*-BN dielectric, close to the high-resistance Dirac point [13]. As we reach a threshold voltage (~ –1.4 V for Cycle 1 in figure 2a), the carrier density begins to increase at a significantly higher rate, suggesting the intercalation of Li ions into the heterostructure. At the same threshold voltage, we observe a spike in the sheet resistance of the device, consistent with a decrease in the graphene mobility as charged Li ions moves into the device. The exact threshold voltage value varies somewhat from device to device.

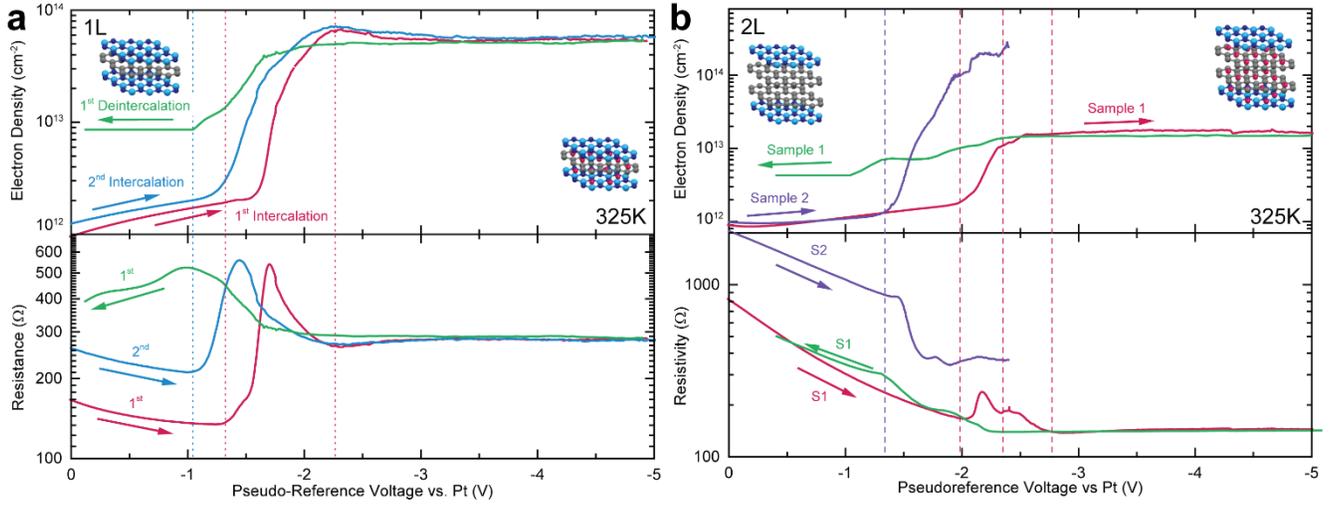

*Figure 2:* Measured electron density and sample resistivity in intercalating and de-intercalating graphene heterostructures fabricated with **(a)** 1-layer graphene and **(b)** 2-layer graphene. In both cases, on the first intercalation cycle (red), electron density increases linearly at low potentials, while resistance decreases, consistent with electrostatic gating across the h-BN. Once the electrochemical potential exceeds a certain threshold, electron density suddenly increases by about an order of magnitude and then saturates, while at the same time resistance spikes. This suggests the onset of the electrochemical reaction. When the potential is swept back towards zero (blue), the sample de-intercalates at a slightly lower voltage, but follows the same general trend in reverse. In **(a)** green shows a second intercalation cycle on the same device, where the same trend is observed, but with lower threshold voltage. **(b)** For the 2LG case, data from a second device is shown instead (purple), showing an intercalation run to the highest achieved electron density in 2LG.

On the reverse scan, we observe a concordant steep decrease of the charge carrier density associated with a somewhat broader maximum in the resistance of the device. We note that de-intercalation happens at a lower voltage than the initial intercalation threshold voltage. Immediately after de-intercalation, our measurements indicate the graphene heterostructures contain more electrons than the pristine device, suggesting that some Li ions remain between the sheets initially. After holding the device at 0 V overnight (> 12 hours), the device returned to its fully de-intercalated state where the residual charge density becomes ~ $10^{12}$ cm$^{-2}$. Subsequently, a second intercalation sweep in the same geometry reveals a response very similar to the first intercalation. Overall, the reversibility of the reaction demonstrates that our measurements are not a result of a sudden delamination of the van der Waals heterostructure, which would result in a permanent increase of electrostatic gating efficiency at all potentials.

Our Hall voltammetry demonstrate that the reversible electrochemical intercalation and deintercalation of Li ions in the interface between graphene and *h*-BN crystals is possible. For 2LG samples, the intercalated Li ions can insert either in the graphene/graphene interface or the graphene/*h*-BN interfaces. Raman spectroscopy is a useful probe to investigate the distribution of intercalation in few-layer-graphene intercalations [15, 16]. Specifically, the G-peak of graphene (wavenumber near 1600 cm$^{-1}$) is a good indicator of the charge density in the graphene basal plane [16].

We performed *in-situ* Raman spectroelectrochemistry to further confirm that Li ions can intercalate the graphene/*h*-BN interface. Figure 3 shows the evolution of the Raman G and 2D peaks of graphene as a function of applied cell potential for both 1LG and 2LG samples. The 1LG structure is once again very instructive. As the cell potential increases, the graphene G peak sharpens and shifts at low voltages due to increasing carrier density in graphene. A fit of the G peak position in figure 3b suggests that the carrier density increases approximately linearly with the applied voltage (so that the Fermi energy increases approximately as $V_{eg}^{1/2}$), consistent with electrostatic gating through the *h*-BN layer. As the cell voltage approaches the threshold, the G peak rapidly blue-shifts, and both G and 2D peaks subsequently disappear, while the corresponding *h*-BN peak [17] at 1370 cm$^{-1}$ remains visible. This suggests that the graphene layer is sufficiently doped by the Li intercalant as to be Pauli-blocked, while the *h*-BN layer remains unaffected. Comparing with graphite literature [16], Pauli-blocking and the vanishing of the graphite Raman peaks is a signature of stage one intercalation in bulk Li intercalated graphite LiC$_6$. This observation further suggests that Li is coming into direct contact with the graphene, and supports the notion of graphene/*h*-BN interface intercalation. Using the excitation laser wavelength ($\lambda$ = 532 nm) as a lower bound for the Fermi energy, the graphene was doped to $E_F$ > 1.16 *eV*, corresponding a charge density of 9.9 × $10^{13}$ cm$^{-2}$.

In both the 1LG and 2LG heterostructures, the Raman spectra do not exhibit an appreciable D peak at 1350 cm$^{-1}$, which is commonly associated with chemical disorder or damage of the in-plane covalent graphene bonds [18]. The absence of Raman D peak during the intercalation/deintercalation process suggests that the Li ions are inserted into the graphene/*h*BN interface without damaging the in-plane bonds between the carbon atoms, leaving the graphene lattice itself intact. Furthermore, Raman spectra taken on the same device through multiple cycles (fig. 3c) demonstrate the reversibility of the Raman behavior. The graphene Raman peaks reappear when we reach a threshold voltage during the reverse cell potential scan. A second intercalation cycle produces similar Raman spectra as the first cycle with the G peak shifting as a function of gate potential.

The 2LG structure undergoes two forms of intercalation: at the graphene/graphene interface and at the graphene/*h*-BN interface. We observe behavior in Raman that cannot be ascribed solely to intercalation of the graphene/graphene interface (Fig. 3d). In bulk graphite, stage 2 behavior (LiC$_{12}$), which denotes that only one side of each graphene layer is occupied by the intercalant, is characterized by a broadened, blue-shifted, but nevertheless present G peak [16]. In our 2LG device, the Raman G and 2D peaks both disappear at high voltage (while the *h*-BN D mode remains unchanged), indicating doping beyond the C$_6$LiC$_6$ stage (> 3 x $10^{14}$ cm$^{-2}$) for each layer of graphene.

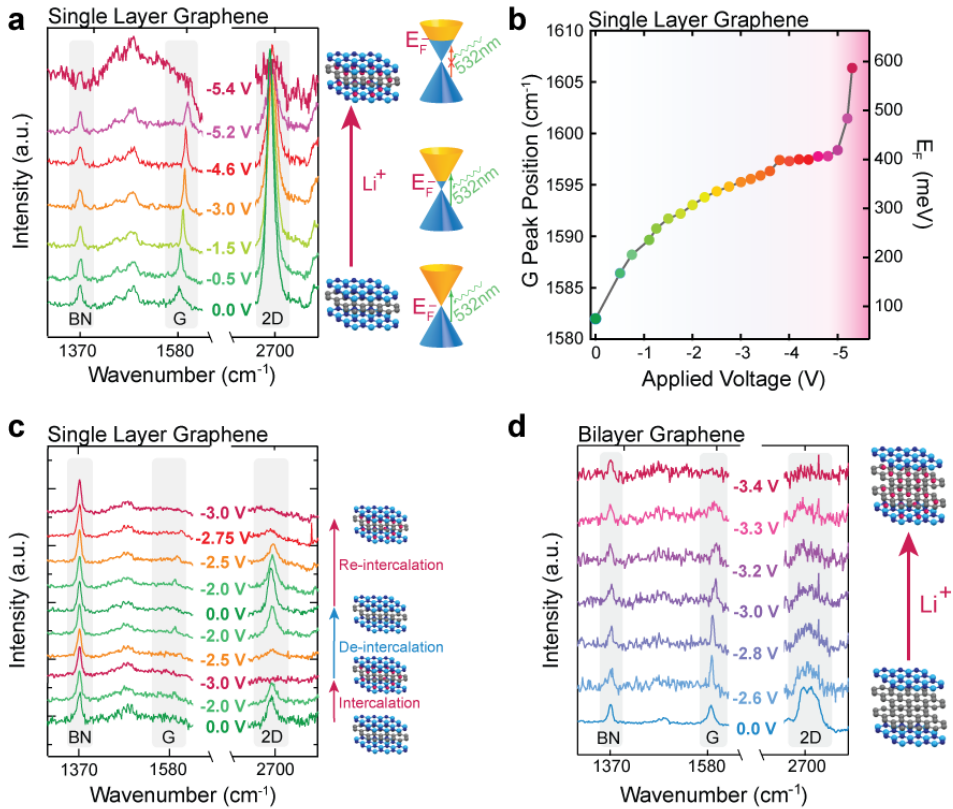

*Figure 3: (a, c, d)* Representative Raman spectra demonstrating the evolution of the G and 2D peaks with applied potential while the h-BN peak remains constant for 1LG (a, c) and 2LG (d) encapsulated in h-BN. Electrolyte fluorescence background is subtracted from all Raman spectra, and the spectra are offset for clarity. Note the gradual blue-shift in graphene G peak. Beyond the threshold intercalation voltage, all graphene spectroscopic signatures disappear. Dirac cones show, schematically, the doping of the graphene crystal and subsequent Pauli blocking. The relative intensities of the peaks differ from optical interference effects [14] arising form differences in the heterostructure thickness, due to the choice of h-BN crystals. *(b)* G-peak position measured from spectra shown in (a) as a function of applied voltage for 1LG encapsulated in h-BN. Note the sudden change in G peak position immediately prior to its disappearance, showing that the graphene had suddenly intercalated after reaching the threshold voltage.

The cycling Hall measurement described above was performed at slightly elevated temperature (325 K or 52 °C) to enable simultaneous monitoring of the carrier density during the intercalation process. We also performed low temperature transport characterization of intercalated samples. For this experiment, the intercalated device was cooled quickly (at a cooling rate ~10 K/min), while holding the applied cell potential. It is known that Li ions become immobile in the PEO matrix when cooled below the polymer glass transition temperature of $T_g$ ~200 K. Below $T_g$, all electrochemical processes are completely frozen out and no deintercalation processes occur even when the cell potential is brought back to 0 V. This allows us to create stable intercalated vdW heterostructures.

With the samples frozen at a pre-defined electrochemical potential, we measure the Hall voltage as a function of the applied magnetic field. We find that the electron density of the 1LG heterostructures saturates around a maximum of ~7 × $10^{13}$ cm$^{-2}$, although the exact value varies between devices and maximum applied electrochemical potentials. Based on density functional theory (DFT) calculations suggesting that each intercalating Li ion donates 0.88 electron charge to the conducting graphene system [10], we calculate a Li:C stoichiometric ratio of approximately 1:60, about one-tenth of the electron density of stage 1 Li intercalated bulk graphite. This ratio is smaller yet consistent with our DFT calculations suggesting an upper bound of Li:C stoichiometry ~1:20 in *h*BN/graphene/*h*BN sandwich structure (see Supplementary Fig. S1). Overall this data indicates that the graphene/*h*-BN interface is far less amenable to hosting Li atoms than two neighboring graphene planes.

The intercalation process injects a large number of Li ions into the graphene/*h*-BN heterostructure. At the Li densities achieved here, the Li atoms are likely to be distributed randomly relative to the graphene lattice, creating scattering sites for the conducting electrons in graphene. Correspondingly, we observe a large decrease in mobility after intercalation, from ~200,000 cm$^2$/(Vs) readily achieved in clean 1LG devices to approximately ~1500 cm$^2$/(Vs). Despite this relatively low mobility value, we observe Shubnikov-de Haas (SdH) oscillation of magnetoresistance $R_{xx}$ as a function of applied magnetic field $B$ at 1.8 K (fig. 4a). The observed SdH oscillations in intercalated 1LG heterostructure exhibit a single period oscillation in $B^{-1}$ (inset fig. 4a). Assuming spin and valley degeneracy of 1LG, this density estimated from SdH oscillation is 4 x $10^{13}$ cm$^{-2}$, in agreement with the density measured from the Hall measurement.

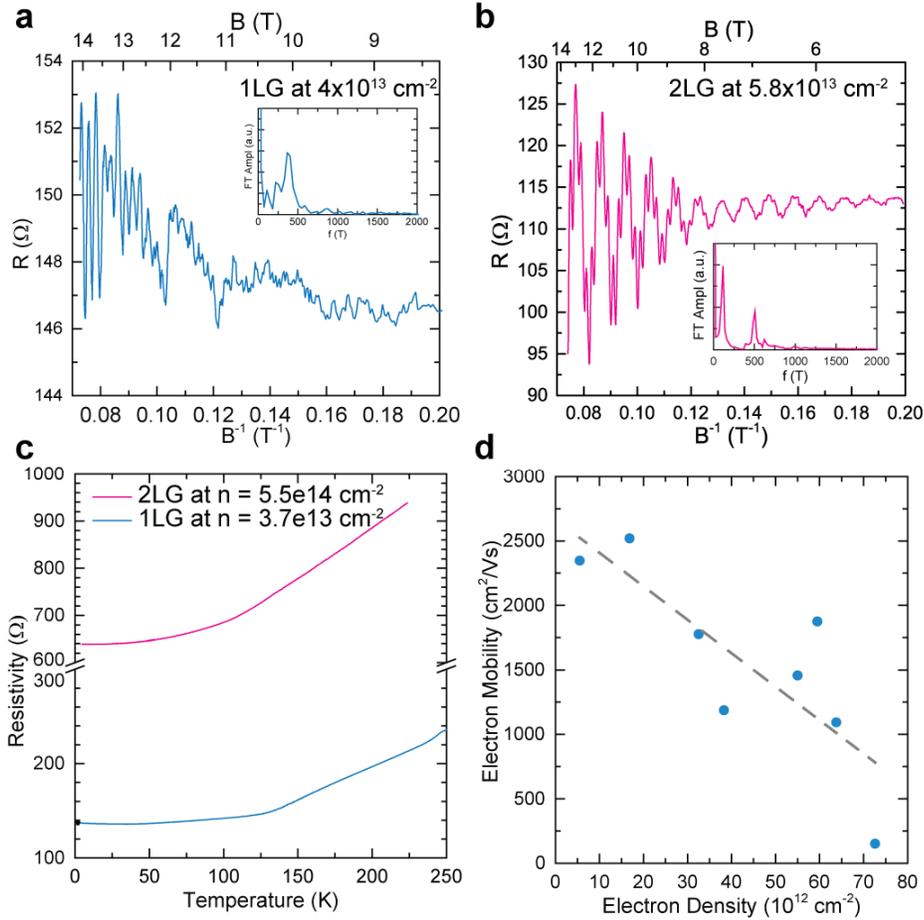

*Figure 4: (a) Measured $R_{xx}$ as a function of $B^{-1}$ in a 1LG device with Hall carrier density $4.0 \times 10^{13}$ cm$^{-2}$ at 1.8 K. Periodic SdH oscillations are visible. The inset shows Fourier transform of SdH, indicating a peak corresponding to the carrier density $(3.66 \pm .08) \times 10^{13}$ cm$^{-2}$. (b) $R_{xx}$ as a function of $B^{-1}$ in a 2LG device with Hall carrier density $5.8 \times 10^{13}$ cm$^{-2}$ at 1.8 K. The inset shows Fourier transform of SdH, indicating two peaks corresponding to the carrier densities $(1.07\pm01) \times 10^{13}$ cm$^{-2}$ and $(4.84 \pm.01) \times 10^{13}$ cm$^{-2}$. (c) Resistivity of intercalated 1LG (blue) and 2LG (red) heterostructures at cryogenic temperatures, showing metallic behavior. (d) Electron mobility vs. electron density from multiple intercalated 1LG devices, cooled to below 10 K. Line is a guide to the eye showing a general decrease of mobility after increasing intercalation.*

Similar measurements performed on 2LG heterostructures revealed the maximum carrier density up to $5.5 \times 10^{14}$ cm$^{-2}$. By assuming that the carrier density of the graphene/$h$-BN interfaces in 2LG matches that of the 1LG heterostructure ($7 \times 10^{13}$ cm$^{-2}$), we can estimate an electron density in the graphene/graphene interface of $4.8 \times 10^{14}$ cm$^{-2}$. This data indicates that the graphene-graphene interface can host significantly more Li ions than the $h$-BN/graphene interface in agreement with our DFT calculation (see SI).

SdH oscillations are also visible in the 2LG devices, but only if the electron density is kept low ($< 6 \times 10^{13}$ cm$^{-2}$) by keeping the cell potential close to the threshold before the cooldown. As shown in figure 4b, the SdH oscillations show two distinct frequencies, unlike the 1LG case, corresponding to conduction in two separate bands. Similar two SdH oscillation frequency have been observed in electrolytically gated bilayer graphene samples [19], suggesting different populations in the lower and higher subbands of 2LG. In our experiment, the total carrier density in both bands matches with the value derived from Hall data.

Finally, we discuss the temperature dependent transport in the intercalated compound. For this measurement, we slowly warm up the samples (warming up rate ~5 K/min) under magnetic field and measure $R_{xx}$ and Hall resistance $R_{xy}$ of the samples to obtain the density and mobility of the samples at different temperatures. Figure 4c shows that $R_{xx}$ increases monotonically for 1LG and 2LG as temperature increases, indicating metallic behaviors of both samples. We note that both mobility and density typically remain approximately constant below 200 K. However, when the intercalated devices are warmed above the polymer $T_g$, chemical reactions resume as a function of the cell potential. We note that the electron mobility decreases significantly at higher temperatures and at higher doping. At the highest carrier density of $5.5 \times 10^{14}$ cm$^{-2}$ for 2LG, the electron mobility decreases to ~18 cm$^2$/(Vs) when T < 100 K. An overview of the mobilities attained in multiple 1LG devices intercalated to different densities is presented in Figure 4d. We observe a general trend in decreasing mobility with increasing intercalation density, consistent with increasing inter-valley scattering due to increasing numbers of Li ions associated with the graphene layer.

In summary, we have demonstrated the electrochemical intercalation of Li into graphene encapsulated between $h$-BN layers. Passsivation of the device components (graphene surface and electrodes) prevents electrochemical side reactions that lead to the modification of the sample surface. Our device platform allows for *in-situ* characterization of the doping level and electrical transport properties as the intecalation progresses. Using our novel Hall-voltametry method, we can not only very precisely monitor the intercalation through Hall effect, but we can also intercalate the galleries in the graphene/*h*-BN interface. The effect of intercalation into vdW heterostructure is most prominent in the 1LG case, where gating effects are not enough to explain the observed high doping levels in both the Raman and transport data. Our technique enables the engineering of novel vdW heterostructures with diverse functionality and applications.

## AUTHOR INFORMATION

**Corresponding Author**

* Email: pkim@physics.harvard.edu

**Present Addresses**

D.K.E now at ICFO - The Institute of Photonic Sciences, Mediterranean Technology Park, Av. Carl Friedrich Gauss 3, 08860 Castelldefels (Barcelona), Spain

J.R. now at Mork Family Department of Chemical Engineering and Materials Science, University of Southern California, Los Angeles, California 90089, United States

Y.G. now at Department of Chemistry, Northwestern University, Evanston, Illinois 60208, United States

**Notes**

The authors declare no competing financial interests.

## ACKNOWLEDGMENT

We thank E. Kaxiras and S. N. Shirodkar for fruitful discussions. The work is supported by the Harvard collaboration was supported by the Science and Technology Center for Integrated Quantum Materials, NSF Grant No. DMR-1231319 and the Nano Material Technology Development Program through the National Research Foundation of Korea (NRF) funded by the Ministry of Science, ICT and Future Planning (2012M3A7B4049966). P.K. acknowledges support from the ARO MURI Award No. W911NF14-0247. X.R. acknowledges support from Air Force Office of Scientific Research under AFOSR Award No. FA9550-14-1-0381. G. E. acknowledges support by the SRC-NRI Hans J. Coufal Fellowship and the Columbia Optics and Quantum Electronics NSF IGERT (DGE-1069240). S. Y. F. Z. acknowledges support by the NSERC PostGraduate Scholarship. S. H. and K.-A. M. acknowledge support by the Priority Research Center Program (2010-0020207) and Basic Science Research Program (2017R1A2B2010123) through the National Research Foundation of Korea (NRF). K.W. and T.T. acknowledge support from the Elemental Strategy Initiative conducted by the MEXT, Japan and JSPS KAKENHI Grant Numbers JP26248061, JP15K21722 and JP25106006. Device fabrication was performed at the Harvard Center for Nanoscale Systems, NSF award number ECS-0335765.

# Supplementary Information

## 1. Detailed Computational Methods

To obtain the stoichiometric ratio of Li:C for Li intercalations in graphene/*h*-BN van der Waals (vdW) heterostructures, we have carried out density functional theory (DFT) calculations within generalized gradient approximation (GGA) for exchange-correlation (*xc*) functionals [1,2], implemented in the Vienna *ab initio* simulation package (VASP) [3,4]. The kinetic energy cut-off for the planewave basis set is set to 400 eV, and electron-ion interactions are represented by the projector augmented wave (PAW) potentials [5,6]. For the vdW corrections, we adopt the Grimme's DFT-D3 method [7] based on a semi-empirical GGA-type theory. For the calculations, we consider 1-layer graphene (1LG) and 2-layer graphene (2LG) sandwiched between *h*-BN layers, and ($\sqrt{3} \times \sqrt{3}$), (3 × 3), ($2\sqrt{3} \times 2\sqrt{3}$), ($3\sqrt{3} \times 3\sqrt{3}$), and (6× 6) surface unit cells of graphene and *h*-BN are used with 1.88% of lattice mismatch between them. Atomic coordinates are fully optimized until the Hellmann-Feynman forces are less than 0.03 eV/Å.

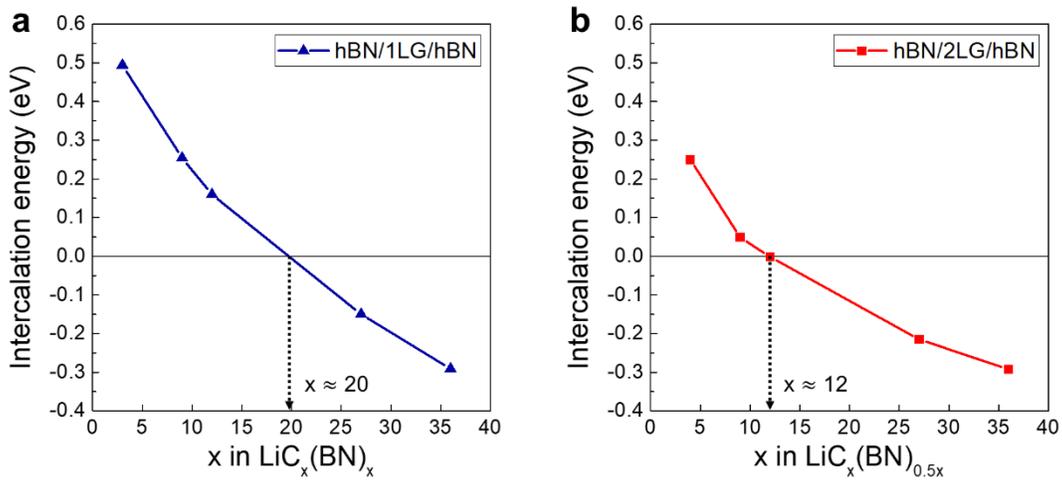

**Figure S1:** Intercalation energy of Li atoms in van der Waals (vdW) heterostructures depending on the Li:C stoichiometric ratio for (a) 1-layer graphene (1LG) and (b) 2-layer graphene (2LG) sandwiched between hBN layers.

## 2. Density Functional Theory (DFT) Calculations for Li Intercalation in Graphene/h-BN van der Waals Heterostructures

To obtain the stoichiometric ratio of Li:C for Li intercalations in graphene/*h*-BN van der Waals (vdW) heterostructures, we have performed density functional theory (DFT) calculations. In Figure S1, we show the intercalation energy of Li atoms depending on the stoichiometric ratio of Li:C. Here, the intercalation energy ($E_{inter}$) of Li atoms is defined by

$$E_{inter} = [E_{(G/hBN)+Li} - E_{G/hBN} - nE_{Li}]/n ,$$

where $E_{(G/hBN)+Li}$ and $E_{G/hBN}$ are total energies of the Li intercalated graphene/*h*-BN heterostructure and the graphene/*h*-BN heterostructure, respectively. $E_{Li}$ is the bulk energy of a Li atom and *n* is the number of Li atoms. If the intercalation energy becomes negative ($E_{inter} < 0$), Li prefers the intercalation into the heterostructure. For the Li intercalations, we consider two types of vdW heterstructures which are 1-layer graphene (1LG) and 2-layer graphene (2LG) sandwiched between *h*-BN layers. From Figure S1(a) and (b), upper bounds of Li:C stoichiometry ratio are found to be about 1:20 and 1:12 in *h*-BN/1LG/*h*-BN and *h*-BN/2LG/*h*-BN sandwiched structures, respectively, which indicates the larger upper bound of Li:C stoichiometry ratio in *h*-BN/2LG/*h*-BN. Such behavior can be expected due to the fact that Li atoms can be intercalated in both graphene-*h*-BN and graphene-graphene interfaces of *h*-BN/2LG/*h*-BN structure, while it can be intercalated in only graphene-h-BN interfaces of h-BN/1LG/h-BN structure. For comparison, we identify the upper bound of Li:C stoichiometry ratio to be about 1:26 for Li atoms in h-BN/graphene interface. The Li:C ratio is found to be much smaller than that in the graphene-graphene interface. These findings imply that such graphene-graphene interface plays a significant role in Li intercalation of the vdW heterostructures. Therefore, it is concluded that Li atoms can be more intercalated in h-BN/2LG/h-BN than h-BN/1LG/h-BN structures.

## *3. Detailed Experimental Methods*

Few layer graphene and h-BN were prepared by mechanical exfoliation of kish graphite and bulk h-BN directly onto Si substrates with 285 nm of $SiO_2$. H-BN/graphene/h-BN sandwich structures were fabricated using a poly(propylene oxide) polymer based dry transfer technique [8]. Subsequently, the heterostructures were etched into a Hall bar shape in an inductively coupled plasma reactor (Surface Technology Systems MPX/LPX ICP RIE), using $O_2$ and $CHF_3$ plasmas and poly(methyl methacrylate) (PMMA) resist mask. Cr/Pd/Au (2/15/60 nm thick) electrical contacts to the graphene crystal, as well as Ti/Pt counter and reference electrodes, were fabricated using standard electron-beam lithography (Raith 150 electron beam writer) using PMMA resist, physical vapor deposition, and lift-off process in acetone. Finally, a SU-8 protective coating was applied over the metal contacts using a standard electron-beam lithography at low dose (~2 $\mu C/cm^{-2}$ at 30 keV). This serves to suppress parasitic reactions on the gold contacts.

Lithium bis(trifluoromethane)sulfonimide (LiTFSI) (Sigma-Aldrich) and BHT inhibitor free poly(ethylene oxide) (PEO) ($M_w$ = 100,000 g/mol) (Alfa Aesar) were purchased commercially and were dried under reduced pressure at 180 °C and 55 °C respectively. Acetonitrile was transferred from activated 4 Å molecular sieves. Li salts and PEO were mixed under Ar atmosphere to achieve a molar ratio of 38 : 1 (PEO monomer units : LiTFSI) and dissolved in acetonitrile. The electrolyte solution is drop-casted using a micropipette over pre-fabricated graphene devices in an argon glovebox, with oxygen and water contaminants kept below 0.1 ppm, and the ensuing film was dried at 320K under vacuum overnight to remove trace solvents. Throughout the process, the utmost care was made in keeping the electrolyte and device away from contaminants in order to minimize any parasitic reactions.

A Keithley 2400 SourceMeter provided the electrical potential to the Pt counter-electrode, while an Agilent 34401A digital multimeter set to > 10 GΩ input impedance read back the pseudo-reference Pt electrode potential. The graphene working electrode was kept grounded. In order to perform the simultaneous resistance and Hall effect measurements, a Stanford Research Systems (SRS) SR-830 lock-in amplifier supplied 1 µA of current at 17 Hz flowing across the Hall bar device. The graphene devices were < 1 kΩ in resistance, therefore the AC contribution to the graphene working electrode potential was less than 1 mV, significantly less than the electrochemical intercalation energy scale.

The Hall effect measurements were performed in a Quantum Design Physical Properties Measurement System (PPMS) cryostat, or a Cryo Industries of America (CIA) RC102-CFM microscopy cryostat system custom-modified for electrical transport measurements, and with a CIA 5T air-bore superconducting magnet. Crucially, to keep the electrolyte free of air contamination, the sample was either mounted to the sealed cryostat directly inside an argon glovebox (for the microscopy cryostat), or the sample package (a ceramic DIP-16 chip carrier) was sealed using a glass cover slip glued to the sealing ring using Dow Corning vacuum grease. This cover was able to protect a sensitive chemical indicator for oxygen from air (under ambient conditions) for about three days.

Raman spectra were collected on a Renishaw inVia spectrometer at 532 nm laser wavelength. Laser spot size was focused to ~2 µm. The samples were loaded into the microscopy cryostat with a thin 0.5 mm optical window and heated to 325K under high vacuum (~$10^{-5}$ mBar) for 12 hours before initiating the electrochemical reaction.

􀀀